\documentclass[a4paper,14pt]{extarticle}
\usepackage{cmap}
\usepackage[cp1251]{inputenc}
\usepackage[english,russian]{babel}
\usepackage{braket,amssymb,amsmath,array,commath,mathtext,wrapfig,tikz}
\usepackage[arrows=pgf-filled,version=3]{mhchem}
\usepackage{bm}
\usepackage{amsthm,amsfonts,amscd}
\usepackage{graphicx}
\usepackage{cite}
\usepackage{lscape}
\usepackage{geometry}
\usepackage{indentfirst}
\usepackage{afterpage,soul,ulem,dcolumn,changebar,longtable,hhline,
	multirow,makecell}
\usepackage[small,bf]{caption}
\makeatletter
\renewcommand{\@biblabel}[1]{#1.\hfil}
\makeatother

\begin{document}
\renewcommand{\refname}{References}
\renewcommand{\figurename}{Fig}
\begin{center}
\textit{On the temperature of hairy black holes}
\end{center}
	\begin{center}
 \textbf{Vitalii Vertogradov}$^{1}$,
 \textbf{Dmitriy Kudryavcev}$^{2}$\\
	\end{center}

	\begin{center}
		$^{1}$ \quad 
  Physics department, Herzen state Pedagogical University of Russia,
		
		48 Moika Emb., Saint Petersburg 191186, Russia
		
		SPb branch of SAO RAS, 65 Pulkovskoe Rd, Saint Petersburg 196140, Russia
		
		vdvertogradov@gmail.com \\
  $^{2}$ \quad 
  Physics department, Herzen state Pedagogical University of Russia,
		
		48 Moika Emb., Saint Petersburg 191186, Russia
 \\ kudryavtsiev33@gmail.com
	\end{center}

\textbf{Abstract:} The gravitational decoupling method represents an extremely useful tool to obtain new solutions of the Einstein equations through minimal geometrical deformations. In this paper, we consider the hairy charged black hole obtained by the gravitational decoupling and calculate their Hawking temperature in order to compare it with the case when the hairs are ignored. We have found out that the hairs, under some conditions of black hole parameters,  affect the Hawking temperature and can increase it. We  have also found out that the black hole temperature, in hairy case, doesn't depend on the electric charge.

\textbf{Key words}: Hairy black hole, Hawking temperature, Charged black hole, gravitational decoupling.

\section{Introduction}

Recent paper~\cite{bib:grib} showed the possibility of phase transition near the event horizon of a black hole. However, the critical Hawking temperature~\cite{bib:hawking} near horizon, at which the phase transition can happen,  is reached in the vicinity of the event horizon and the radius of the region, where this effect is possible, is negligible. When one considers the particle collisions, the situation becomes better because the center-of-mass energy can grow unboundly in some processes~\cite{bib:bsw, bib:vaidyacol, bib:pavlov, bib:zaslavanti}. The region near the event horizon where phase transition can happen is much bigger in comparison to the Hawking temperature~\cite{bib:grib}. However, one can try to increase the effect caused by the Hawking temperature by considering the modification of the standard black hole solution and how these modifications affect the black hole temperature.

There is a famous theorem in black hole theory which states that black hole doesn't have hairs i.e. black hole can have only three charges - a mass $M$, angular momentum $j$ and electric charge $Q$. However, it was shown that a black hole can have soft hair~\cite{bib:hawking_hair}. Recently, it was understood that one can obtain a hairy black hole by using the gravitational decoupling method~\cite{bib:gd1, bib:gd2, bib:gd3}. Obtaining  new analytical solutions of the Einstein equations is an extremely hard task in most cases. One can solve these equations, for example, for the spherical symmetry and the perfect fluid as the source of the gravitation. However, if~
we consider the more realistic case when the perfect fluid is coupled to another matter, it is nearly impossible to obtain the analytical solution. The gravitational decoupling through minimal geometrical deformation shows the possibility of decoupling two gravitational sources. One can write the energy-momentum tensor as :
\begin{equation} \label{eq:thefirst}
T_{ik}=\tilde{T}_{ik}+\alpha\Theta_{ik} \,.
\end{equation}
where $\tilde{T}_{ik}$ is the energy-momentum tensor of the perfect
fluid and $\alpha$ is the coupling constant to the energy-momentum tensor
$\Theta_{ik}$. 
It is possible to solve Einstein's field equations for a gravitational
a source whose energy-momentum tensor is expressed as
\eqref{eq:thefirst} by solving Einstein's field equations for each
component $\tilde{T}_{ik}$ and $\Theta_{ik}$ separately. Then, by~a
straightforward superposition of the two solutions, we obtain the
complete solution corresponding to the source $T_{ik}$. Since the Einstein equations are not linear, this method represents an useful method of analysis of the solution. It is an especially important tool when one  faces the realistic cases i.e. the stars and collapsing objects which interior matter is far from an ideal perfect fluid.

By applying this method a new modification of Schwarzschild black hole has been obtained~\cite{bib:bh1, bib:bh2}. These black hole solutions satisfy the strong and dominant energy conditions in the whole region  from the event horizon up to infinity. All these solutions have been obtained by small deformations of the Schwarzschild vacuum. The more realistic case i.e. the deformation of dynamical background has been obtained in~\cite{bib:vermax}

In this paper, we consider the Hawking temperature  for hairy charged Reissner-Nordstrom  black hole obtained by gravitational decoupling through minimal geometrical deformation. And how it deviates from the black hole when the hairs are ignored. All these models represent a black hole supported by a non-linear electrodynamics. 

This paper is organized as follows: in sec. 2 we briefly discuss two methods of obtaining the Hawking temperature for general spherically-symmetric black hole. In sec. 3 explcitly calculate the Hawking temperature for Reissner-Nordstrom black hole. In sec.4 we consider three models of  hairy black hole, their temperature and compare these results with no-hair solution. Section 5 is the conclusion and discussion of future research plan.
The system of units $c=G=1$ will be used throughout the paper. Also, we adopt the signature $-\,, +\,, +\,, +$.

\section{Black hole thermodynamics}

In this section we review the basic concepts related to the black hole thermodynamics. For more useful and thorough  discussion on this subject, one can see, for example, the review~\cite{bib:review}. We consider spherically-symmetric line element in the form:
\begin{equation} \label{eq:metric}
ds^2=-fdt^2+f^{-1}dr^2+r^2 d\Omega^2 \,.
\end{equation}
Where a lapse function $f=f(r)$ depends upon radial coordinate $r$, $d\Omega^2=d\theta^2+\sin^2\theta d\varphi^2$ is the metric on unit two-sphere. To describe black holes, it is convenient to write \eqref{eq:metric} in the form:
\begin{equation} \label{eq:shape}
ds^2=-\left( 1-\frac{b(r)}{r}\right) dt^2+\left(1-\frac{b(r)}{r} \right)^{-1}dr^2+r^2d\Omega^2 \,.
\end{equation}
Here, we can refer to the function $b(r)$ as the shape function which specifies  the shape of the spatial slice. In the limit $\lim\limits_{r\to infty} b(r)$ the shape function can be interpreted as asymptotic mass  $2M$. We assume the asymptotic flatness for all models considered in this paper. The metric \eqref{eq:shape} has horizons  at $b(r_h)=r_h$. This equation might have several roots but only the outermost horizon is in main interest and we will consider only this one. We interested in the case that $\forall r>r_h \rightarrow b(r)<r$  and $\frac{db(r)}{dr}|_{r=r_h}<1$. The case $\frac{b(r)}{dr}|_{r=r_h}=1$ corresponds to an extremal black hole for which the Hawking temperature is zero~\cite{bib:wisser} and this case won't be considered within this paper.
The Hawking temperature is given by:
\begin{equation} \label{eq:hawking}
k_bT_h=\frac{\hbar \varkappa}{2\pi} \,.
\end{equation}
Where $k_b$ is the Boltzmann constant, $T_h$  is Hawking temperature and $\varkappa$ is the surface gravity which for the metric \eqref{eq:metric} has the form:
\begin{equation}
\varkappa = \lim\limits_{r\rightarrow r_h}\frac{1}{2}\frac{df}{dr} \,.
\end{equation}
Substituting a lapse function in the form \eqref{eq:shape}, one obtains:
\begin{equation}
\varkappa=\frac{1}{2r_h}\left [1-b'(r_h) \right] \,.
\end{equation}
where a dash corresponds to the derivative with respect to the radial coordinate $r$. 

Another way obtaining the Hawking temperature is Euclidean signature techniques.  By the formal transformation to the imaginary time $t \rightarrow it$ gives:
\begin{equation} \label{eq:evklid}
ds^2=fdt^2+f^{-1}dr^2+r^2d\Omega^2 \,.
\end{equation}
Again, we are interested in the outermost horizon at $r=r_h$ and discarding the whole $r<r_h$ region. Furthermore, we assume the lapse function $f$ in the form given by \eqref{eq:shape} and also demand that a black hole is not an extremal one. Taylor expand near the horizon gives:
\begin{equation} \label{eq:series}
1-\frac{b(r)}{r} \approx \left [1-b'(r_h)\right ] \frac{r-r_h}{r_h} \,.
\end{equation}
Substituting \eqref{eq:series} into \eqref{eq:evklid}, one obtains:
\begin{equation} \label{eq:aprox}
ds^2\approx \frac{(1-b'(r_h))(r-r_h)}{r_h}dt^2+\left[ \frac{(1-b'(r_h))(r-r_h)}{r_h} \right]^{-1}dr^2+r^2_hd\Omega^2 \,.
\end{equation}
By introducing a new variable $R$:
\begin{equation} \label{eq:transformation}
dR=\left [\frac{(1-b'(r_h))(r-r_h)}{r_h} \right ]^{-\frac{1}{2}} dr \,,
\end{equation}
one can transform the metric \eqref{eq:aprox} to obtain:
\begin{equation} \label{eq:period}
ds^2\approx \frac{\left[1-b'(r_h)\right]^2}{4r_h^2}R^2 dt^2+dR^2+r_h^2d\Omega^2 \,.
\end{equation}
The $(t, R)$ part of this metric is similar to a flat two-plane in polar coordinates, with imaginary time $t$ serving as the angular coordinate. In order to avoid a conical singularity, one should demand that $\frac{1-b'(r_h)}{2r_h}t$ has a period $2\pi$ i.e. $t$ has a period $\beta$ which is given by:
\begin{equation} \label{eq:per}
\beta =\frac{4\pi r_h}{1-b'(r_h)} \,.
\end{equation}
According to~\cite{bib:17} this imaginary time $t$ is interpreted as  the existence of  the thermal bath of a temperature $k_bT_h=\frac{\hbar}{\beta}$ which explicitly gives:
\begin{equation}
k_bT_h=\frac{\hbar}{4\pi r_h}\left (1-b'(r_h)\right) \,,
\end{equation}
which coinsides with \eqref{eq:hawking}.

\section{The Hawking temperature of \\ Reissner-Nordstrom black hole }

In this section we will apply the method described in the previous section to Reissner-Nordstrom solution which describes the charged static black hole. The result of this section is well-known and can be found, for example, in ~\cite{bib:wisser, bib:pois, bib:new}. We rederive these results only in order to compare them with hairy black holes.

From here and further in the paper  we will use the system of units $k_b=\hbar=1$. The lapse function $f$ \eqref{eq:metric} in Reissner-Nordstrom case is given by:
\begin{equation} \label{eq:nordstrom}
f(r)=1-\frac{2m}{r}+\frac{Q^2}{r^2} \,.
\end{equation}

The shape function $b(r)$ \eqref{eq:shape}  and its derivative are given by:
\begin{equation}
\begin{split}
b(r)=2M-\frac{Q^2}{r} \,, \\
b'(r)=\frac{Q^2}{r^2} \,.
\end{split}
\end{equation}
The Reissner-Nordstrom black hole has two horizons which are located at:
\begin{equation}
r_{\pm}=M\pm \sqrt{M^2-Q^2} \,.
\end{equation}
Here, as we stated above, we are interested only in the outermost horizon, so that $r_h=r_+$. The cases $Q^2>M^2$ and $Q^2=M^2$, which correspond to a naked singularity and an extremal Reissner-Nordstrom black hole, won't be considered within this paper. 

The surface gravity $\varkappa$ at the horizon is given by:
\begin{equation}
\varkappa= \frac{1}{r_h} \left( 1-\frac{Q^2}{r_h^2} \right ) \,. 
\end{equation}
 And the Hawking temperature is:
\begin{equation}
T_h=\frac{1}{4\pi r_h}\left ( 1-\frac{Q^2}{r_h^2} \right) \,.
\end{equation}

Note, that in the extremal case $M^2=Q^2$, the horizon is located at $r_h=M$ which gives $1-\frac{Q^2}{r_h^2}=0 \rightarrow T_h=0$.

\section{Thermodynamics of hairy black holes}

In recent paper~\cite{bib:bh1}, new solution which describe the exterior geometry of hairy black holes, have been introduced by using the gravitational decoupling method. In this section, we will calculate the Hawking temperature and compare results with usual Reissner-Nordstrom black hole in order to find out how primary hairs affect the Hawking temperature.

\subsection{The model 1}

This model can be interpreted as a black hole supported by a non-linear electrodynamics. Other two models can be obtained from this model by defining the electric charge $Q$ as a function of Schwarzschild mass $M$ and primary hairs $\alpha$ and $l_0=\alpha l$. 

The lapse function $f$ for the first model is given by:
\begin{equation} \label{eq:lapse1}
f(r)=1-\frac{2\mu}{r}+\frac{Q^2}{r^2}-\frac{\alpha \left(\mu-l_0 /2 \right) e^{-r/(\mu-l_0/2)}}{r} \,.
\end{equation}
Here $\mu = M+l_0/2$, $M$ is the Schwarzschild mass, $\alpha$ is the coupling constant and $l_0=\alpha l$ is the primary hair, $Q$ can be interpreted as the electric charge of a black hole. The influence of the geodesic motion of primary hairs has been studied in the paper~\cite{bib:geod}. The influence of these parameters for hairy Schwarzschild black hole has been done in~\cite{bib:thermo}. The Schwarzschild solution is the limit of $\alpha \rightarrow 0$. 

The event horizon equation is given by $f(r_h)=0$ which can be solved with respect to $l_0$ to give:
\begin{equation} \label{eq:horizon1}
l_0 =r_H-2M+\frac{Q^2}{r_H}-\alpha M e^{-r_H /M} \,.
\end{equation}

We have several restrictions on the parameters. First of all, one should realize that like in pure Reissner-Nordstrom case, one should impose the following condition on them as $M$ and charge $Q$  - $M^2 \geq Q^2$ in order to avoid  a naked singularity in usual Reissner-Nordstrom case. However, the condition $M^2=Q^2$ is not forbidden because in this case the hairy black hole \eqref{eq:lapse1} is not an extremal one. Also, this model satisfies the dominant energy condition only when $r \geq 2M$, so we don't consider the region $0 \leq r \leq 2M$ and demand $r_h\geq 2M$ to satisfy the energy condition. The fulfilling of the dominant energy condition  also imposes the following restrictions on parameters $Q$ and $l$:
\begin{equation} \label{eq:conditions}
\begin{split}
|Q|\geq\alpha \frac{M^2}{e^2} \,, \\
l\geq \frac{M}{e^2} \,.
\end{split}
\end{equation}
 
The hawking temperature \eqref{eq:hawking}  $T_h=\frac{\varkappa}{4\pi}$ is given by in terms of the surface gravity $\varkappa$ which for this model reads:
\begin{equation}
2\varkappa=1+\frac{2M}{r_h^2}-\frac{2 Q^2}{r_h^3}+\frac{\alpha r_h e^{-r_h/(M-l_0/2)} +\alpha \left(M-l_0 /2 \right) e^{-r_h/(M-l_0/2)}}{r_h^2} \,.
\end{equation}
So the Hawking temperature is:
\begin{equation} \label{eq:temperature1}
T_h=\frac{1}{4 \pi} \abs{1+\frac{2M}{r_h^2}-\frac{2 Q^2}{r_h^3}+\frac{\alpha r_h e^{-r_h/(M-l_0/2)} +\alpha \left(M-l_0 /2 \right) e^{-r_h/(M-l_0/2)}}{r_h^2}}
\end{equation}

Fig.1 is plotted for $M=1\,, \alpha=0.5\,, Q=0.9$ This figure shows the dependence the primary hair $l_0$ from the horizon location (blue curve). The horizontal red line corresponds to  $l_0= \alpha \frac{M}{e^2}$. We see that  at $r_h\geq 2.061$ the dominant energy condition is always held.
\begin{figure}[ht!]
    \centering
    \includegraphics{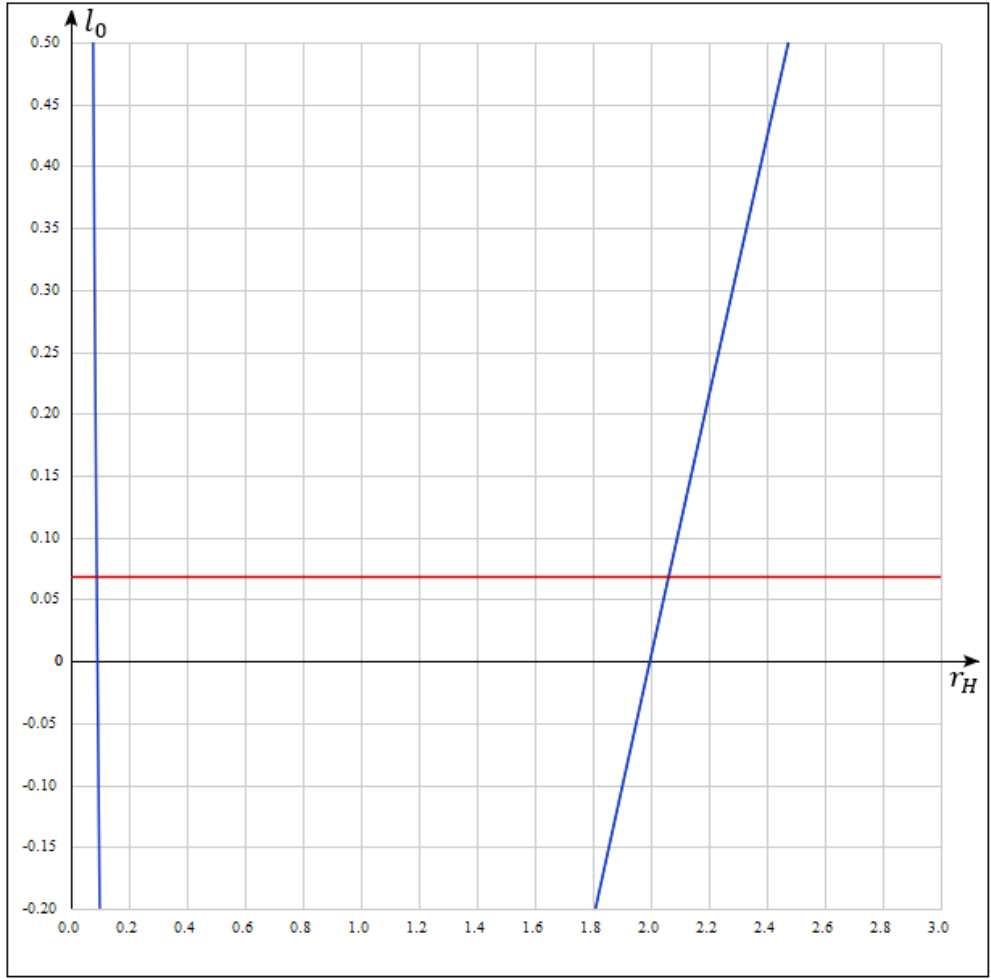}
    \caption{The event horizon function}
    \label{fig:figure11}
\end{figure}
Fig. 2 shows the dependence the Hawking temperature $T_h$ from the horizon location. Three horizontal lines correspond the Hawking temperature of usual Reissner-Nordstrom black hole. We have picked up the following parameters for horizontal lines: $M=1$, $Q=0.5$ (red line); $Q=0.9$ (green line) and $Q=0.99$ (orange line). In this model the Hawking temperature  of hairy black hole doesn't depend on the electric charge $Q$ and is the function only a mass $M$, coupling constant $\alpha=0.5$ and horizon location $r_h$ i.e. $T_h\equiv T_h (M, \alpha, r_h)$ (the corresponding curve is in blue). We can easily see that at $r_h  \in (2.061\,, 0.212)$ for $Q=0.5$; $r_h\in (2.061\,, 2.481)$ for $Q=0.9$ and $r_h \in (2.061\,, 4.583)$ for $Q=0.99$ the Hawking temperature is higher than for usual Reissner-Nordstrom black hole. It means that the phase transition, which can happen near the event horizon, in these cases, can be fairer than in no-hair black hole. One can also see that when one considers an extremal Reissner-Nordstrom black hole, then the Hawking temperature is absent but in the hairy case it is not zero. So, we can conclude that for the first model the primary hairs can increase the Hawking temperature in the comparison the usual Reissner-Nordstrom case.
\begin{figure}[ht!]
    \centering
    \includegraphics{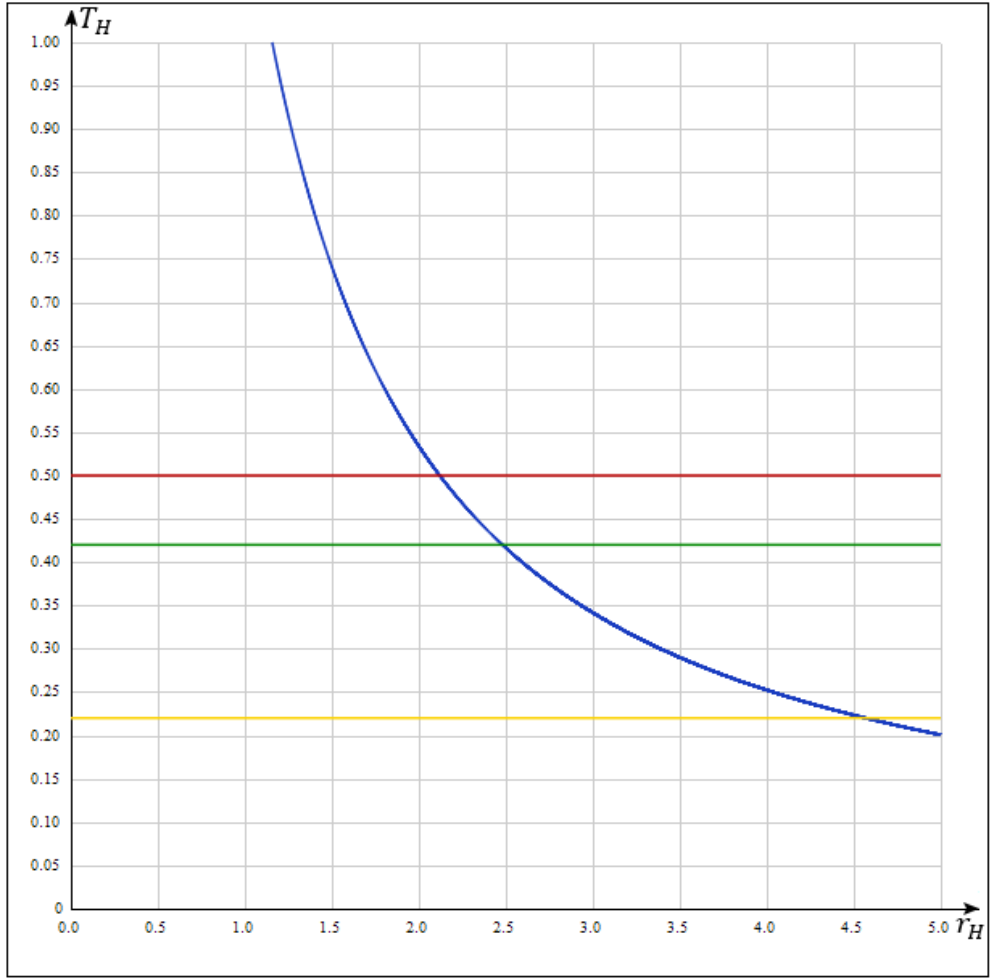}
    \caption{The Hawking temperature. Model 1.}
    \label{fig:figure2}
\end{figure}
\subsection{Model 2}

As we stated in the previous subsection, the ensuing models differ from the first one by appropriate choice of the function $Q$. In this model $Q$ is given by:
\begin{equation}
Q^2=L_0M\left (2+\alpha e^{-l_0/M} \right) \,.
\end{equation}
Substituting this into \eqref{eq:lapse1}, one obtains:
\begin{equation} \label{eq:lapse2}
f=1-\frac{2M+l_0}{r}+\frac{2 l_0 M}{r^2}-\frac{\alpha M e^{-r/M}}{r^2} \left(r-l_0 e^{\frac{r-l_0}{M}} \right)
\end{equation}
For this choice of the charge function the horizon equation is:
\begin{equation}
r_h=l_0 \,.
\end{equation}

Like in the previous subsection, the Hawking temperature is given in terms of the first derivative of the lapse function $f'$, which in this case reads:
\begin{equation}
\begin{split}
f'=\frac{2M+l_0}{r_h^2}-\frac{4 l_0 M}{r_h^3}+\frac{\alpha e^{-r_h/M} \left(r_h-2M\right)}{r_h^3}\left(r_h-l_0 e^\frac{r_h-l_0}{M}\right)+ \\
+\frac{\alpha M e^{-r_h/M}}{r_h^2}
\left(\frac{l_0}{M} e^\frac{r_h-l_0}{M}-1 \right) \,.
\end{split}
\end{equation}

In this model, the dominant energy condition is always held for $r\geq 2$. Again, like in the first model, the Hawking temperature, after substituting $l_0$, doesn't depend on the electric charge of a hairy black hole. 
\begin{figure}[ht!]
    \centering
    \includegraphics{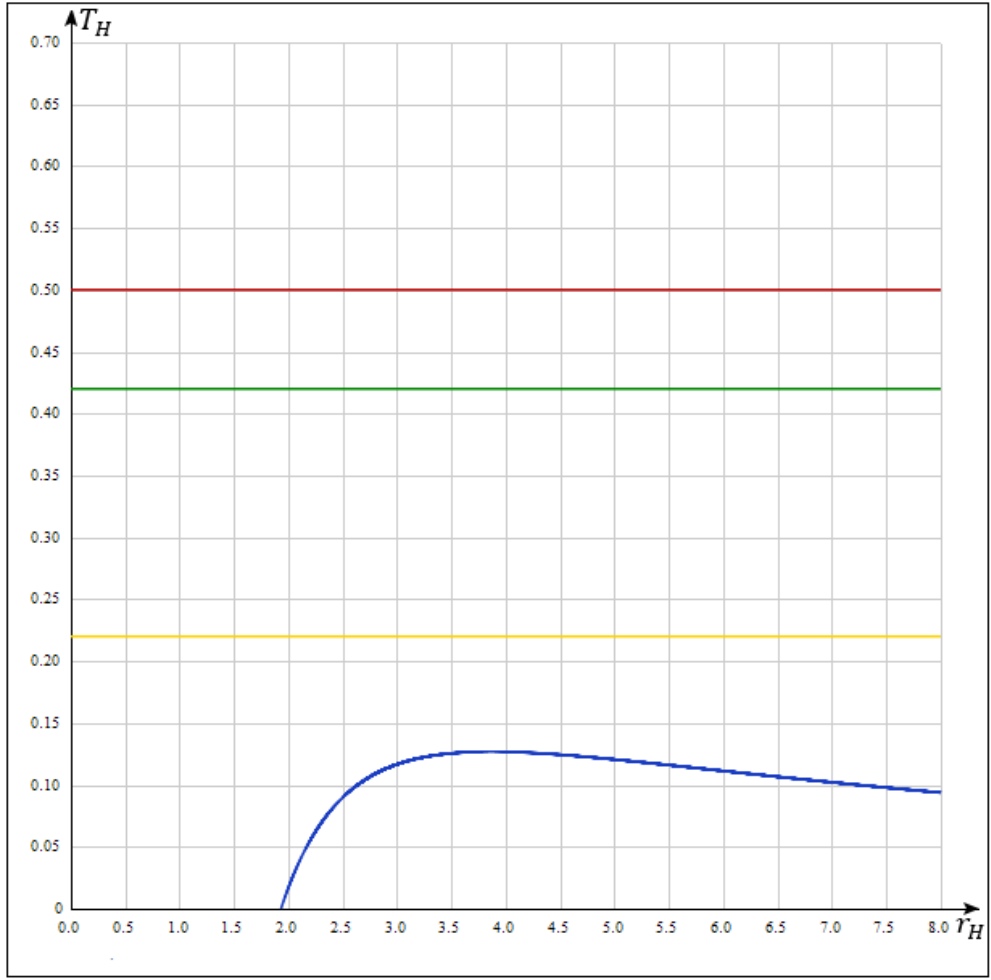}
    \caption{The Hawking temperature. Model 2.}
    \label{fig:figure12}
\end{figure}
Fig. 3 shows how the Hawking temperature $T_H$ depends on the event horizon location $r_h$. The choice of the parameters is like in the first model. This figure differs from the fig.2 only the blue curve, which corresponds to the Hawking temperature of a hairy black hole. From the figure, one can see that the temperature of a hairy black hole is always less than in  a no-hair case. The only exception is the extremal Reissner-Nordstrom black hole. So we can conclude that one can consider the bigger region, where phase transition can happen, only for an extremal Reissner-Nordstrom black hole. In other cases the region is less than in no-hair case.

\subsection{Model 3}

For the last model, which we consider within this paper, the charge function is given by:
\begin{equation}
Q^2 =\alpha M \left(2M+l_0 \right) e^{-\frac{2M+l_0}{M}} \,.
\end{equation}
Substituting it into the lapse function $f$ \eqref{eq:lapse1}, gives us:
\begin{equation}
f = 1-\frac{2M+l_0}{r} - \frac{\alpha M}{r^2} e^{-r/M}\left(r-\left(2M+l_0\right) e^{\frac{r-2M-l_0}{M}}\right) \,.
\end{equation}
The horizon in this model is located at:
\begin{equation}
r_H=2M+l_0 \,.
\end{equation}
And the first derivative of the lapse function, which gives the main contribution to the Hawking temperature, reads:
\begin{equation}
\begin{split}
f' = \frac{2M+l_0}{r_h^2}+\frac{2\alpha M}{r_h^3} e^{-r_h/M}\left(r_h-\left(2M+l_0\right) e^{\frac{r_h-2M-l_0}{M}}\right)+ \\ 
+\frac{\alpha}{r_h^2} e^{-r_h/M} \left(r_h-2M+l_0\right)
e^{\frac{r_h-2M-l_0}{M}}-\frac{\alpha M}{r_h^2} e^{-r_h/M} \left(1-\left(2+l_0/M\right) e^{\frac{r_h-2M+l_0}{M}} \right)
\end{split}
\end{equation}

In this model, the dominant energy condition is held when $r_h>2.067$. It shouldn't be a surprise that for this model the Hawking temperature doesn't again depend on the electric charge of a hairy black hole.

Figure 4 is plotted for the same parameters like in two previous cases and show the Hawking temperature $T_H$ as the function of the event horizon location $r_h$. From the figure, one can see that the hairy black hole temperature    at $r_h \in (2.067\,, 2.074$ for  $Q=0.5$; $r_h \in (2.067, 2.444)$ for $Q=0.9$ and $r_h \in (2.067\,, 4.583)$ for $Q=0.99$ respectively is bigger than in the no-hair Reissner-Nordstrom black hole. This model shows that under the proper choice of parameters the region, where the phase transition can take place, is bigger than in the usual charged black hole.
\begin{figure}[ht!]
    \centering
    \includegraphics{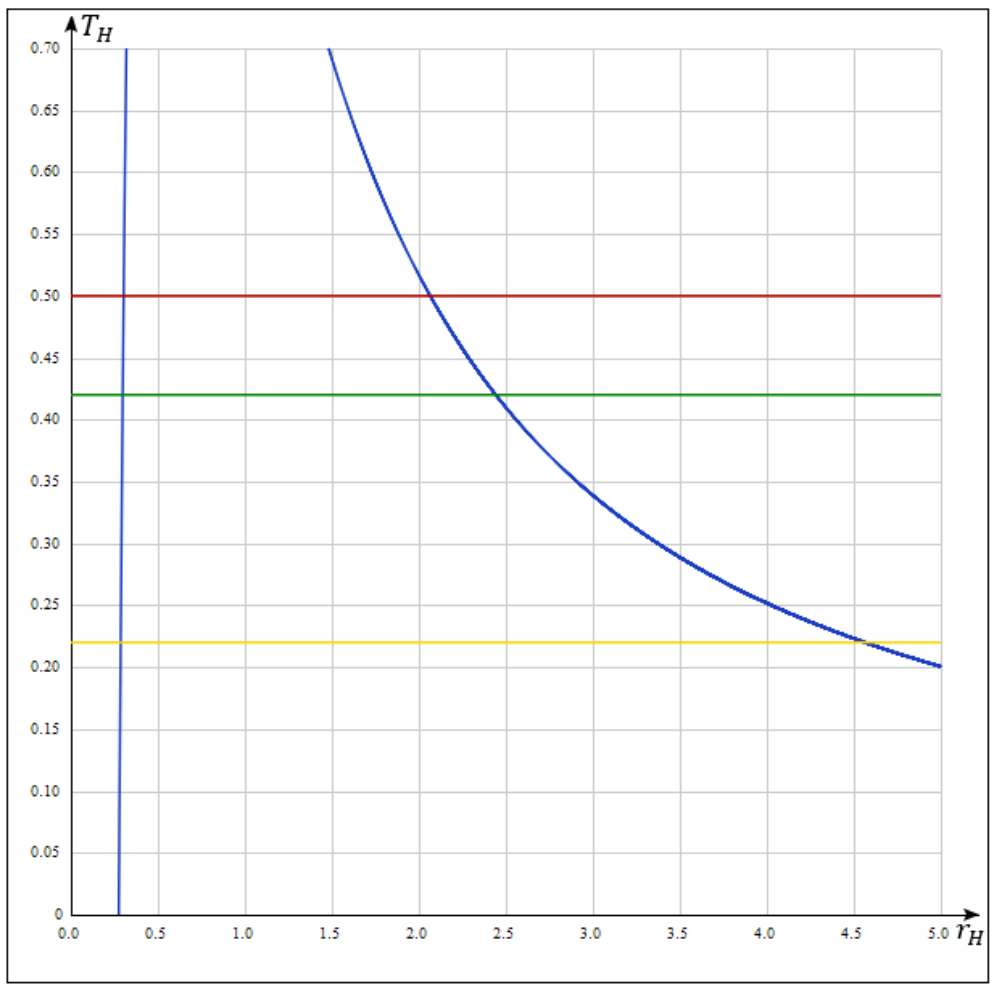}
    \caption{The Hawking temperature. Model 3.}
    \label{fig:figure3}
\end{figure}

\section{Conclusion}

In this paper, we have derived the Hawking temperature of the hairy charged black holes by gravitational decoupling. These black holes can be interpreted as ones supported by a non-linear electrodynamics. The main analytical result obtained within this paper is that the Hawking temperature doesn't depend on the electric charge $Q$ of a hairy black hole. 

We have considered three models and show that the Hawking temperature is not zero in the case of the extremal no-hair Reissner-Nordstrom black hole. However, in the case of the second model, the Hawking temperature of the hairy charged black hole is always less than one in the case non-extremal no-hair charged black hole. For the first and third model, we showed that under certain choice of the primary hairs, one can obtain the black hole temperature higher than in the non-extremal no-hair Reissner-Nordstrom black hole. It means that the region, where the phase transition can take place for the first and the third models, is bigger than the one in usual charged black hole. 

We have left for future research the question about  the entropy, heat capacity and stability of the hairy black hole. The exact value of $r$ at which the phase transition can take place is also the question of the future research.  The particle collisions and corresponding temperature and how hairs of a black hole affect on the process of collision will be considered in the next series of papers.

\textbf{Acknowledgments} Author says thank to grant  22-22-00112 RSF for financial support.

\end{document}